\documentclass[twocolumn]{agujournal2019} 
\usepackage{url} 
\usepackage{soul}
\usepackage{adjustbox}

\draftfalse

\journalname{JGR: Space Physics}

\begin{document}

\title{Geocoronal Solar Wind Charge Exchange Process Associated with the 2006-December-13 Coronal Mass Ejection Event}

\authors{Yu Zhou\affil{1,3}\thanks{Current address: International Center for Quantum-field Measurement Systems for Studies of the Universe and Particles, High Energy Accelerator Research Organization, 1-1 Oho, Tsukuba, Ibaraki 300-3256}, 
Noriko Y. Yamasaki\affil{1,3}, Shin Toriumi\affil{1}, and Kazuhisa Mitsuda\affil{2,3}}

\affiliation{1}{Institute of Space and Astronautical Science, Japan Aerospace Exploration Agency, 3-1-1 Yoshinodai, Sagamihara, Kanagawa 252-5210}
\affiliation{2}{National Astronomical Observatory of Japan, 2-21-1 Osawa, Mitaka, Tokyo 181-8588}
\affiliation{3}{International Center for Quantum-field Measurement Systems for Studies of the Universe and Particles (WPI,QUP), High Energy Accelerator Research Organization, 1-1 Oho, Tsukuba, Ibaraki 300-3256}

\correspondingauthor{Yu Zhou}{zhouyu@post.kek.jp}

\begin{keypoints}
\item We discover a solar wind charge exchange event with Suzaku X-ray satellite 
driven by the 2006-December-13th coronal mass ejection.
\item The SWCX occurence time coincides with the CME magnetic cloud arrival. It is a useful knowledge in space weather forecasting.
\item We modeled the light curve variation of the SWCX. The result indicates that 
the solar-wind flow is anisotropic in the cusp.
\end{keypoints}

\begin{abstract}
We report the discovery of a geocoronal solar wind charge exchange (SWCX) event corresponding to the 
well-known 2006 December 13th coronal mass ejection (CME) event.
Strong evidence for the charge exchange origin of this transient diffuse emission is provided by  
prominent non-thermal emission lines at energies of  $\rm O^{7+}$, $\rm Ne^{9+}$, $\rm Mg^{11+}$, $\rm Si^{12+}$, 
$\rm Si^{13+}$. Especially, a 0.53 keV emission line that most likely arises 
from the $\rm N^{5+}$ $1s^1 5p^1 \to 1s^2$ transition is detected.
Previously, the forecastability of SWCX occurrence with proton flares has been disputed.
In this particular event, we found that the SWCX signal coincided with the arrival of the magnetic cloud inside CME, 
triggered with a time delay after the proton flux fluctuation as the CME shock front passed through the Earth. 
Moreover, a spacecraft orbital modulation in SWCX light curve suggests that the emission arises close to the Earth. 
The line of sight was found to always pass through the northern magnetospheric cusp. The SWCX intensity was high when the 
line of sight passed the dusk side of the cusp, suggesting an azimuthal anisotropy in the flow of solar-wind ions inside the cusp.  
An axisymmetric SWCX emission model is found to underestimate the observed peak intensity by a factor of about 50. 
We suggest this discrepancy is related to the azimuthal anisotropy of the solar-wind flow in the cusp. 

\end{abstract}

\section*{Plain Language Summary}

We discovered compelling observational evidence of the coronal mass ejection (CME) interacting with 
Earth's magnetosphere by the Suzaku satellite. 
The signal arises from a process called charge exchange, in which heavily charged ions 
within the CME interact with neutral hydrogen within Earth's magnetosphere.
This specific process, known as solar wind charge exchange (SWCX), was associated with the CME event 
occurred on 2006 December 13th during the solar minimum.
This event provided us with a unique opportunity to gain insights into the interplanetary consequences of CMEs.
Despite a delay relative to the shock wave associated with the CME, 
the SWCX coincided with the arrival of a magnetic cloud. This suggests that 
solar wind ions were guided along magnetic field lines into a region near Earth called the cusp, 
through a process known as magnetic reconnection. Additionally, we observed that SWCX emissions 
were most pronounced when the satellite's line of sight passed through the dusk region of the cusp. 
This indicates that the CME ejecta may have followed a path along the Parker spiral -- a magnetic field 
pattern twisted by solar rotation -- resulting in the solar wind arriving from a direction distinct from that of the Sun.

\section{Introduction}
\label{section:1}

Charge exchange (CX) process was first observed from comet Hyakutake \cite{Lisse1996}
arising from the interaction between cometary neutrals and solar wind ions \cite{Cravens1997}. 
Nowadays, we have realized that CX event can arise in various scenarios, such as the shock rim 
of supernova remnants \cite{Lallement2009,Katsuda2011,Katsuda2012,Cumbee2014},
the interface between stellar wind and surrounding interstellar medium in starburst galaxies \cite{Pollock2007,Tsuru2007,Liu2011},
merger or accretion shocks inside galaxy clusters \cite{Walker2015,Fabian2011,Gu2015},
and solar wind charge exchange (SWCX) with the planetary neutrals \cite{Snowden2004,Cravens2003,Dennerl2006,Hui2009} 
and interplanetary heliosphere \cite{Cravens2000,Koutroumpa2006,Galeazzi2014,Ringuette2021}.

Geocoronal SWCX results from solar wind ions interacting with the neutrals in Earth's 
atmosphere, leading to transient non-thermal emission lines exceeding the diffuse X-ray background spectrum
\cite{Fujimoto2007,Ezoe2010,Cravens2001, Carter2008, Carter2010, Ezoe2011, Ishikawa2013, Ishi2019, Asakura2021}. 
A significant concern in studying geocoronal SWCX events is forecasting their occurrence 
through solar activity indicators.
Solar wind proton flux has been observed to correlate with 
the charge exchange emission on timescales of about half a day \cite{Fujimoto2007, Carter2010} 
or within two hours \cite{Ezoe2011,Ishikawa2013}. However, \citeA{Ishi2019} demonstrates 
a SWCX event associated with a coronal mass ejection (CME) that lacks significant time coincidence 
with proton flux but instead shows a positive correlation with alpha flux.

Such complication is due to the diverse phenomena triggered by CME eruptions, including 
electromagnetic flares in various energy bands from radio to gamma-ray \cite{Benz2017,DelZanna2018},
solar energetic particles (SEPs) \cite{Desai2016}, highly ionized heavy ions \cite{Lepri2004,Zurbuchen2006}, 
and shock waves \cite{Nindos2011,Verscharen2019} reaching Earth 
with different time delays from the solar surface. The timescales for X-ray brightening are 
short (minutes to hours \cite{Shibata2011}), while impulsive SEP events last a few hours, and large-scale gradual 
SEP events can persist for several days \cite{Reames1999,Desai2016}.
Solar wind (SW), categorized as slow wind (approximately 400 $\rm km\ s^{-1}$), fast wind (around 700-800 $\rm km\ s^{-1}$), 
and coronal mass ejections (CMEs) with speeds up to 2000 $\rm km\ s^{-1}$, approaches Earth over several days or weeks 
\cite{McComas1998,Viall2020,Verscharen2019}. This makes it challenging to establish 
clear correlations between different tracers, especially during frequent solar flares.

Modeling the SWCX spectrum is another challenge, considering uncertainties in the CX cross section, 
ionic composition of the solar wind, relative velocity between solar wind and neutrals, 
and the SWCX interaction geometry with the magnetosphere.

Theoretical approaches to calculate the CX cross section include quantum-mechanical methods \cite{Nolte2012},
atomic-orbital close-coupling \cite{Fritsch&Lin1991}, classical trajectory Monte Carlo \cite{Abrines&Percival1966,Olson&Salop1977}, and multi-channel Landau-Zener method \cite{Janev&Winter1985}. 
However, laboratory measurements of the CX cross section are limited to a small number of ions \cite{Fite1962,Meyer1979,Meyer1985,Crandall1979,Ali2005,Wargelin2005,Greenwood2000,Defay2013,Fogle2014,Beiersdorfer2003,Mullen2017,Mullen2016,Cumbee2018}. 
Consequently, integrated astronomical data analysis software often relies on scaling relations \cite{Gu2016,Olson&Salop1977}
or empirical formulae to estimate these values \cite{Smith2012}.

The SW ionic composition and velocity are determined by factors such as the 
solar flare temperature, plasma equilibrium state at the moment it becomes collisionless, 
acceleration and release mechanisms \cite{Viall2020}, as well as deceleration during propagation 
and the first ionization potential fractionation status under ponderomotive forces induced by magneto-hydrodynamic waves \cite{Laming2015}. 
If SW velocity and CX cross section are accurately constrained, the SWCX spectrum observation can serve as a remote sensor 
for highly ionized ions in the solar wind, complementing in-situ ion composition spectrometers \cite{Gloeckler1998}.

The poorly understood geometry of the geocoronal SWCX interaction is also a critical aspect. 
Studies often assume that solar wind ions fill the magnetosheath, with further penetration 
resisted by magnetic pressure at the magnetopause \cite{Ezoe2011,Carter2010,Ishikawa2013,Ishi2019}. 
However, magnetohydrodynamic (MHD) simulations show prominent X-rays from CX collisions 
between solar wind ions and Earth's exospheric neutrals in the cusp region \cite{Robertson2006}. 
Observations by Suzaku, pointing towards the North pole, indicated that X-ray intensity variation 
is anti-correlated with the geocentric distance from the point where the geomagnetic field 
first opens to space, suggesting SWCX emission can penetrate deeper than the magnetosheath 
near the cusp region \cite{Fujimoto2007}.

In this study, we present a unique case of geocoronal SWCX observed by the Suzaku satellite, 
associated with a well-distinguished source CME occurring on December 13th, 2006. 
The solar environment was near solar minimum at the time, minimizing contamination from 
other events and offering a valuable opportunity to study the space weather consequences of the CME. 
Especially, the satellite's line-of-sight was directed towards the North pole which allows a cusp observation.
Section \ref{section:a.2} provides a comprehensive background description of in-situ measurements for this CME event.
Section \ref{section:a.1} describes the data reduction details for Suzaku observations.
We report the SWCX spectrum detected by Suzaku in section \ref{section:2} and 
explore the time coincidences with interplanetary and Earth magnetic fields in section \ref{section-3}. 
Furthermore, we discuss the SWCX geometry modeling in section \ref{section:4}. Summary and conclusion
are given in section \ref{section:5}.

\begin{figure*}[ht]
\centering
\includegraphics[width=0.6\textwidth]{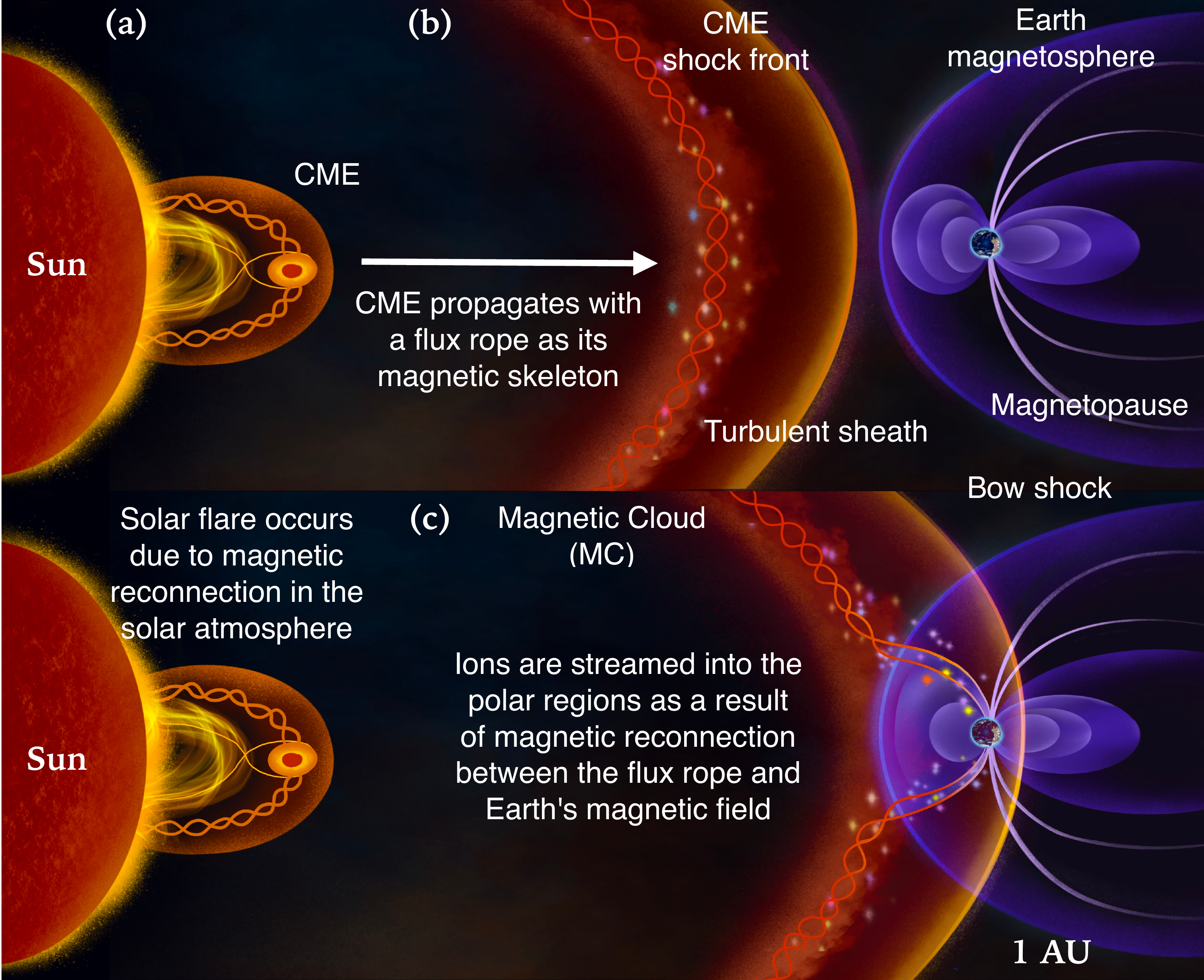}
\caption{A schematic of the CME propagation. (a) Initially, the solar flare occurs and the CME 
is launched near the solar surface as a result of magnetic reconnection, which is observed (in part) 
as abrupt enhancements of soft X-ray and proton flux \cite{Zurbuchen2006,Shibata2011}. 
(b) The CME propagates the interplanetary space (ICME) towards the Earth with the magnetic flux rope as its core. 
The turbulent sheath forms behind the CME shock front and reaches the Earth in advance of the MC. 
(c) As a result of reconnection between the flux rope and Earth's magnetic field, the heavily 
ionized ions are streamed into the magnetosphere, particularly to the polar regions, along the magnetic field lines. 
The artistic drawing is contributed by courtesy of designer Xuelei Chuai.}
\label{fig:schematic}
\end{figure*}

\section{CME Event on 2006 December 13}
\label{section:a.2}

The CME event that occurred on 2006 December 13 was one of the largest halo CMEs 
since the intense solar activity in 2003. This CME was associated with the X3.4-class solar flare, 
emanating from NOAA active region 10930, which was accompanied by the radio bursts, SEP event, 
and MC passage. This flare event has been studied in detail from both observational 
and theoretical points of view, which is summarized in \citeA{HinodeReviewTeam2019} and \citeA{Toriumi2019}.

The coronal observations from the Large Angle Spectroscopic Coronagraph 
and the Extreme ultraviolet Imaging Telescope aboard the Solar and Heliospheric Observatory
shows a strong EUV brightening at 02:30 UT, followed by a dimming and forming into a ring of 
dense material around 03:06 UT \cite{Liu2008}. The CME initial velocity is 1774 $\rm km\ s^{-1}$ near 
the Sun according to the coronagraph height-time observation.
After the abrupt formation, the CME ejecta propagates into the space and its interplanetary CME (ICME) 
consequences, including low-energy and high-energy electrons, protons, MC, were 
measured in-situ by the \textit{STEREO} and \textit{ACE} satellite at 1 AU \cite{Mewaldt2008}. 
At 14:38 UT on December 14, about 36 hours later than the CME outburst, a preceding shock 
passed the spacecraft at an average speed of $\sim$1160 $\rm km\ s^{-1}$ as suggested by 
simultaneous enhancements in the electron flux and magnetic field strength. This speed has 
decreased compared with the initial velocity near the Sun but larger than the shock speed 
as calculated according to mass conservation across the shock at 1 AU ($\sim$1030 $\rm km\ s^{-1}$)\cite{Liu2008}.
Following the preceding shock and magnetic fluctuation period, the detected magnetic field
shows clear rotation, indicating that spacecraft seems to cross the MC interior\cite{Liu2008, Kataoka2009}. 
The shock downstream of the ICME was detected later on 17:23 UT on December 16 with a sharp 
but short duration jump in the proton flux and magnetic field strength. 
The shock was finally detected by the \textit{Ulysses} with a speed of 870 $\rm km\ s^{-1}$ 
at a distance of 2.73 AU to the Sun and to the $117^{\circ}$ east and $74^{\circ}$ south of the Earth at 17:02 UT on December 17 \cite{Richardson2005}. 
But other signatures of ICME, such as enhanced helium abundances, decreased proton temperature, 
or smooth strong magnetic fields were not detected at \textit{Ulysses}. 

\begin{figure*}[ht]
\centering
\includegraphics[width=0.9\textwidth]{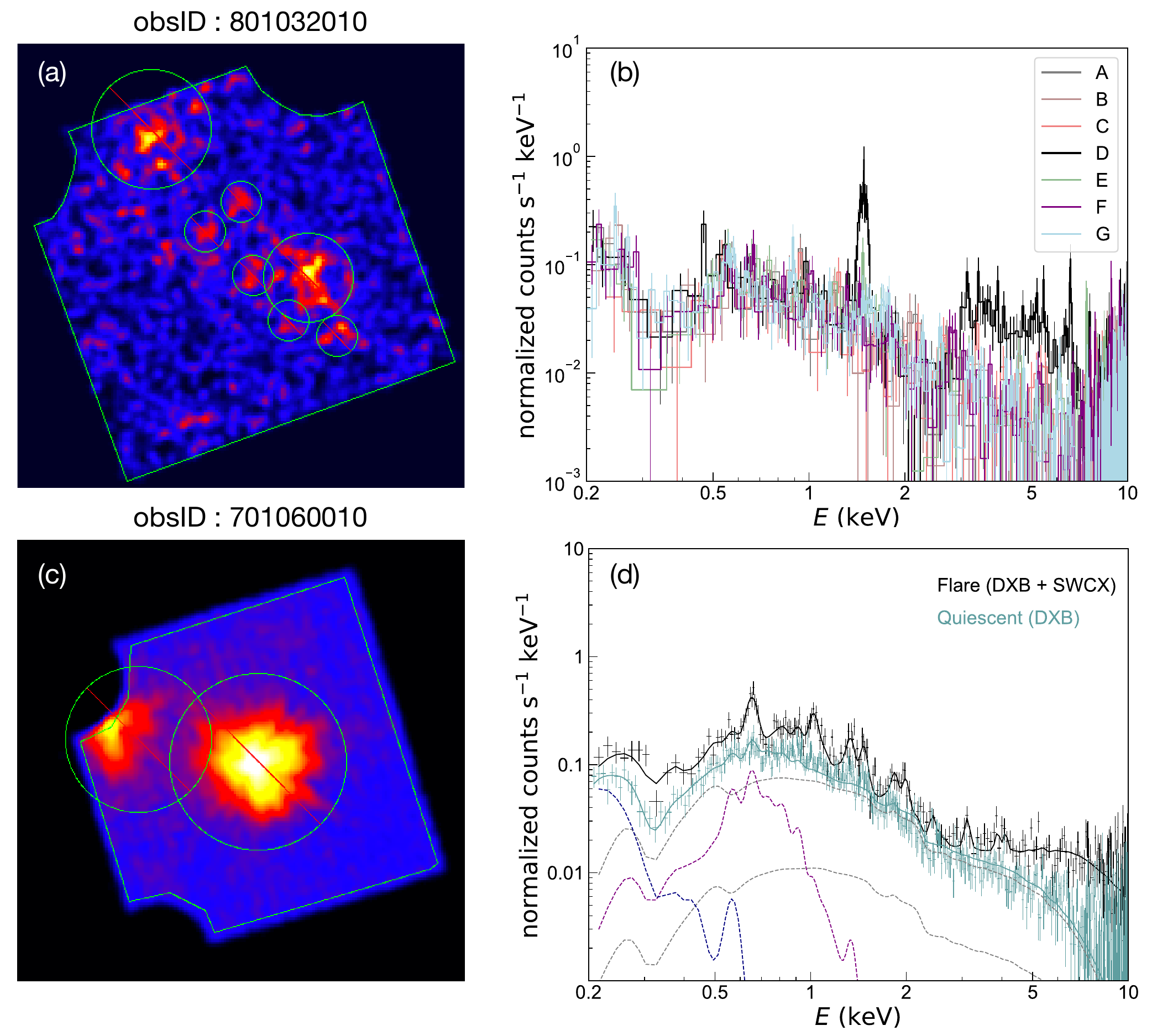}
\caption{(a) and (b) shows the images of the observation 801032010 and 701060010. The green frames 
enclose the CDD regions where the energy spectra were extracted from. 
(c): Energy spectra for the observation 801032010, generated separately for the time intervals [A-G] 
defined from the light curves shown in Figure \ref{fig:region_lc} (b). 
(d): Energy spectra for the observation 701060010 that compare the quiescent and flare time periods defined 
from the light curves shown in Figure \ref{fig:region_lc} (d). Models fitting to  
the diffuse X-ray background break down to three major components: unresolved point sources in the cosmic X-ray background (gray dashed lines),
hot plasma in the galactic gaseous halo (purple dashed line), and warm-hot gas in the local hot bubble (blue dashed line).
}
\label{fig:spec}
\end{figure*}

\begin{figure*}[ht]
\centering
\includegraphics[width=0.9\textwidth]{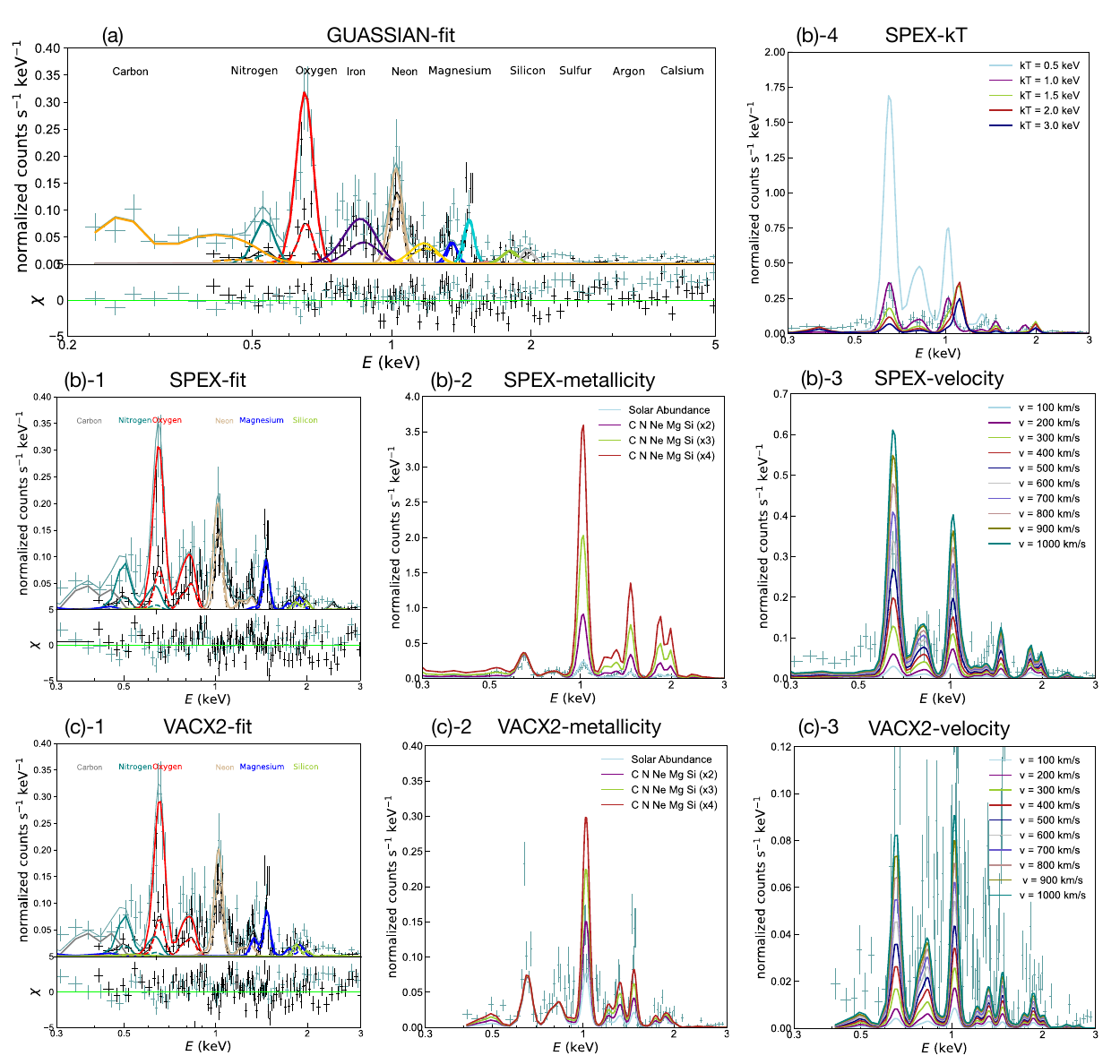}
\caption{(a) Net SWCX spectrum and the fitted Gaussian models for the back-side and front-side illuminated CCDs. 
$blue$ cross: data from XIS1, $black$ cross: merged data from XIS0, XIS2, and XIS3. 
(b)-1 shows the fit of CX model in SPEX. Comparisons of the SPEX-CX model in terms of 
metal abundance, collisional velocity, and plasma temperature are shown in (b)-2, (b)-3, and (b)-4.
Parmeters configuration for (b)-2: ionization temperature $kT = 1.04\rm\ keV$ and collision velocity 
$v = 592.5\rm\ km \ s^{-1}$, and solar abundance for elements other than C, N, Ne, Mg, Si; 
for (b)-3, solar abundance and relative velocity 
$v = 592.5\rm\ km\ s^{-1}$; for (b)-4, solar abundance and ionization temperature $kT = 1.04\rm\ keV$. 
(c)-1 shows the fit of VACX2 model incorporated in XSPEC. Comparisons of the XSPEC-VACX2 model in terms of 
metal abundance and collisional velocity are shown in (c)-2, (c)-3. 
Parameters configuration for (c)-2: plasma temperature $kT = 1.05\rm\ keV$, and solar abundance for elements other than C, N, Ne, Mg, Si;
for (c)-3, solar abundances are set for all elements.}
\label{fig:swcx_gaussian}
\end{figure*}


\begin{figure*}
\centering
\includegraphics[width=0.9\textwidth]{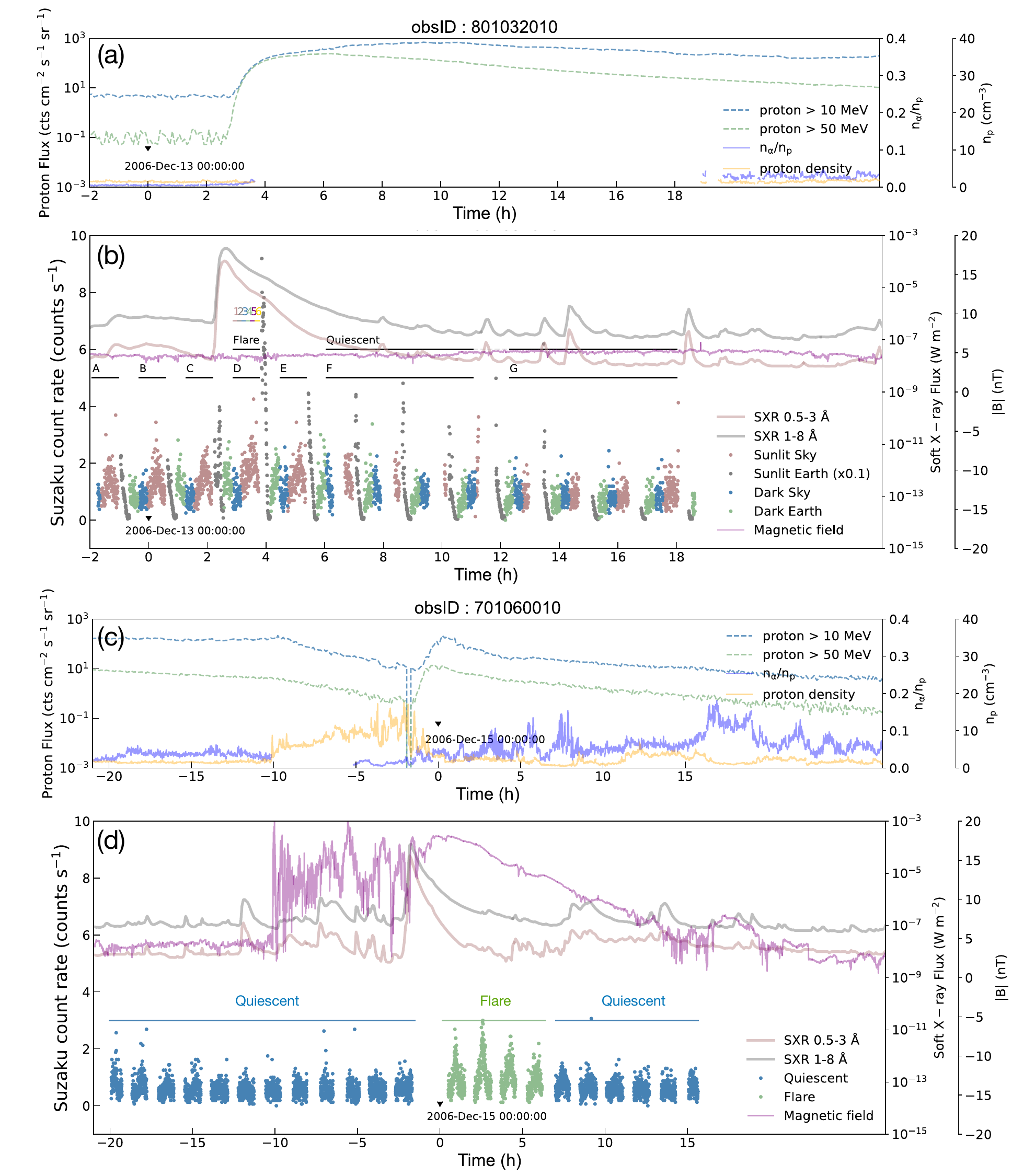}
\caption{The light curves for the Suzaku X-ray background emission and other solar activity tracers.
During the observation 801032010/701060010 period: Proton flux observed by GOES satellite, 
alpha to proton density ratio, and proton density measured by ACE satellite are shown in (a)/(c); 
Count rate of the Suzaku X-ray background emission, soft X-ray flux measured by GOES satellite,
and the magnetic field strength measured by ACE satellite are shown in (b)/(d).
}
\label{fig:region_lc}
\end{figure*}

\begin{figure*}
\centering
\includegraphics[width=0.9\textwidth]{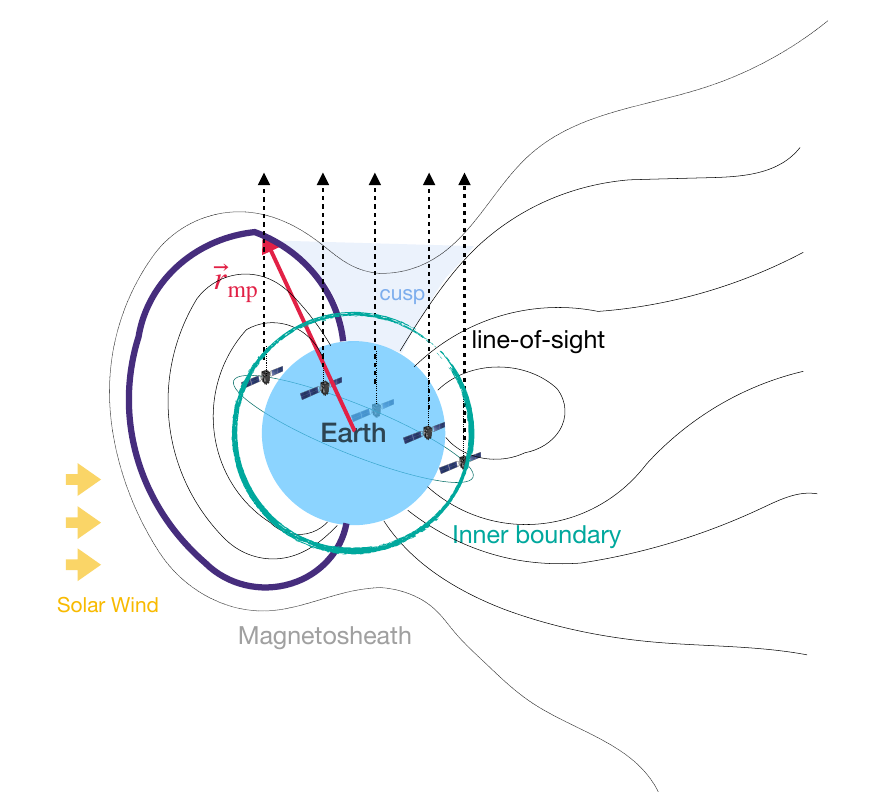}
\caption{A sketch (not drawn to scale) of the Earth, magnetosphere, Suzaku orbit, cusp region, and the definition of 
$\vec{r}_{\rm mp}$ where the line of sight meets the last$-$closed geomagnetic field line.
}
\label{fig:sketch}
\end{figure*}

\begin{figure*}
\centering
\includegraphics[width=0.5\textwidth]{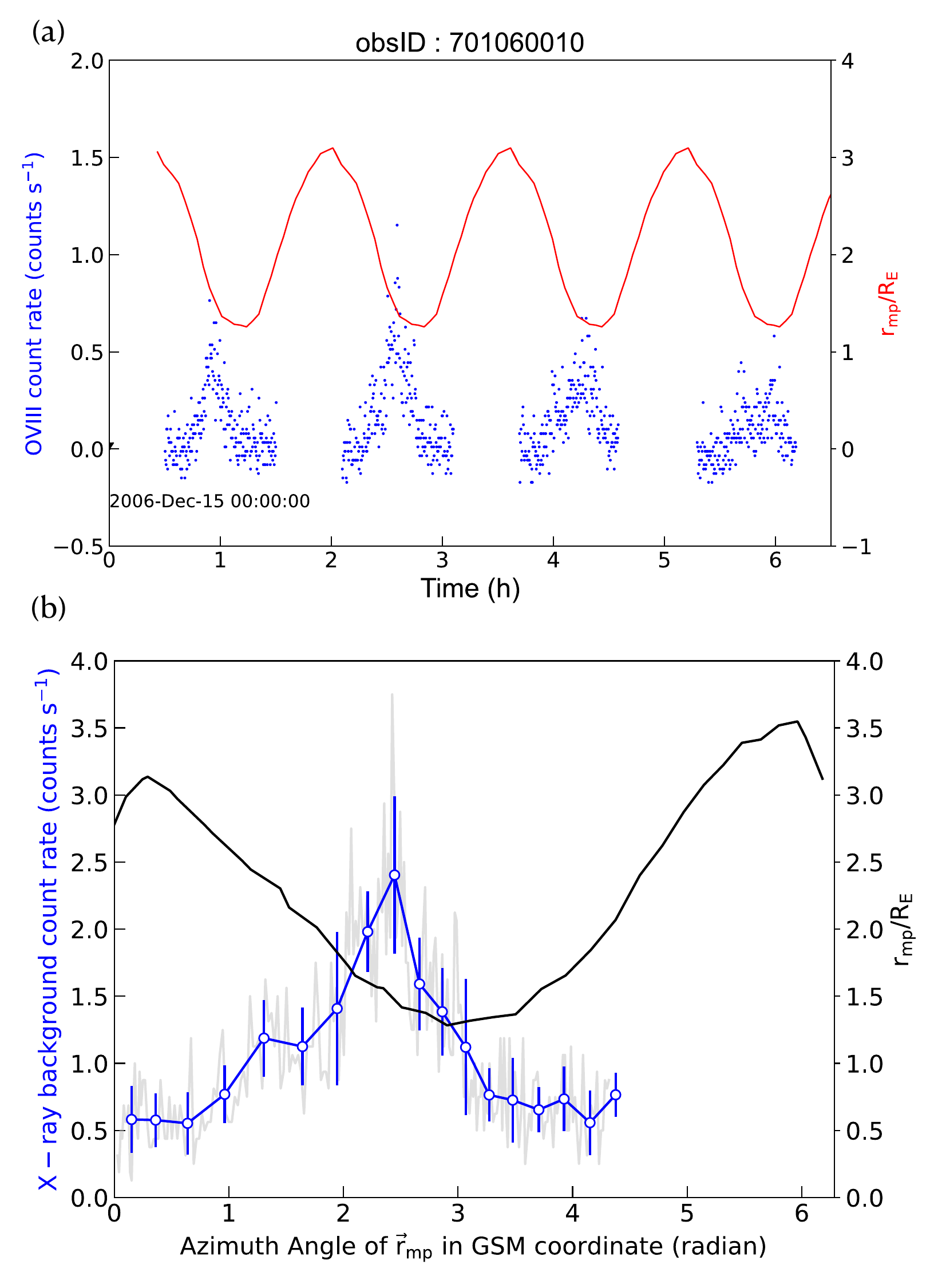}
\caption{(a) Time history of the counting rate for $\rm O^{7+}$ emission line, plotted together with $r_{\rm mp}$, 
the radial distance from the Earth center of the position where the magnetic field is open to out space 
for the first time along the line of site. 
(b) The X-ray intensity and  $r_{\rm mp}$ plotted as functions of the azimuth angle of the $r_{\rm mp}$ point in the GSM coordinate.  The figure shows only for cycle 2 of (a).
}
\label{fig:swcx_geometry}
\end{figure*}

\section{SWCX event captured by Suzaku Observations}
\label{section:a.1}

We analysed the \textit{Suzaku} observations that covers the time period since CME formation to its 
passage across the Earth, including two individual datesets with ObsIDs 801032010 and 701060010. 

\subsection{observation 801032010}
The observation (obsID:801032010) was originally targeted on the galaxy cluster 
Abell 1555, with a few point-sources discernable in the 0.5-2 keV image, as shown in 
Figure \ref{fig:spec} (a). The exposure time of observation, from 2006-Dec-12 22:16 UT 
to 2006-Dec-13 18:38 UT, has covered the CME-accompanied X3.4 flare at 2006-Dec-13 02:14 UT and 
enabled a tracking of detector response since the onset of solar flare.

We select the background region to exclude any contribution from the discrete sources 
and extract the light curve and energy spectrum of the diffuse background. No significant 
flare is observed in the light curve of the dark sky data (selected with the criteria ELV$>$5 \& DYE\_ELV$>$20 \& SUN\_ALT$<$0) 
except for a moderate orbital modulation around the solar flare phase. 
But the sunlit Earth phase data (selected with the criteria ELV$<$0 \& SUN\_ALT$>$0) has shown 
significant enhancement during the solar flare period near 02:14 UT on 2006 December 13, 
which reflects the solar flare photons scattered in the Earth atmosphere.
We further divided the light curve of the flare period into small segments and found an 
abrupt emergence of line emission at $\sim$ 1.5 keV (indicated as [D] period in Figure \ref{fig:region_lc} (b),
and the spectrum shown in Figure \ref{fig:spec} (b). Confirming that the decay time 
constant of the 1.5 keV line is comparable to that of the solar flare X-ray fluxes in 1-8 \AA$\ $energy range, 
rather than that of the proton fluxes detected by GOES satellite, we conclude that this line 
emission is originated from the fluorescent instrumental Al $\rm K\alpha$ 
line (1.486 keV) created by secondary X-ray photons scattered inside the telescope. 

Besides, \textit{Suzaku} detected an excess of blended line emissions in the 3-7 keV energy range 
during the solar flare period, which are probably originated from solar flare photons diffusely reflected
by the Earth atmosphere that were partially collected by the telescope. 
The line emission at 6.62 keV prominent enough to be recognized as highly ionized Fe XXV 
line, which is also seen in the Earth albedo spectrum extracted from the sunlit Earth data, supports this conjecture.

\subsection{observation 701060010}
The observation (obsID:701060010) was pointed towards a Seyfert 1 galaxy, 3C 390.3, with 
two bright sources in the field of view, as shown in the full-band image in Figure \ref{fig:spec} (c).
The timespan of the observation is from 2006-Dec-14 03:25 UT to 2006-Dec-16 03:19 UT, which 
contains the time period when the ICME shock and MC passed by the Earth at around 2006-Dec-14 14:38 UT.
We also noticed an X1.5 flare detected by GOES starting from 22:07 UT on December 14 (see Figure \ref{fig:region_lc} (d)).

We select only the background region to extract the light curve and energy spectrum of the 
diffuse X-ray background emission. As seen in Figure \ref{fig:region_lc} (d), the light curve of Suzaku dark sky
between 2006-Dec-15 0h and 6h shows significant orbital modulation during the flare phase with a higher count rates 
at maxima compared to the quiescent phase. A spectral excess is also clearly observed in the flare phase
in comparison with the spectrum of the quiescent state. The excess emission is dominated 
by individual lines in 0.5-2 keV energy range.

All four \textit{Suzaku} XIS detectors have sensed the transient SWCX excess emission during the 
flare state. We extracted the energy spectra during the flare and quiescent periods for both back-side 
and front-side illuminated CCD detectors.
We merged the spectra for the front-side illuminated sensor XIS0, XIS2, and XIS3 to obtain
a spectrum with higher signal-to-noise ratio. The corresponding response files, ancillary 
files, and non-X-ray background files were also combined using the ASCA \ttfamily{ftool}\normalfont\ command 
\ttfamily{addascaspec}\normalfont. The energy spectra extracted in the quiescent period 
were finally subtracted from the source spectra to exhibit the charge exchange excess for further modelling 
of the net SWCX spectrum. The net SWCX spectra after background subtraction for both types of CCDs 
are shown in Figure \ref{fig:swcx_gaussian} (a).


\section{Geocoronal SWCX energy spectrum}
\label{section:2}
Figure \ref{fig:swcx_gaussian} shows the SWCX energy spectrum fitted with multiple models for both 
front-side and back-side illuminated CCDs.
Table \ref{table:gaussian_fit} shows the fitted centroid energies and flux of the emission lines 
and their probable identifications. 
We found that excess emission lines with fitted centroid energies around 0.652 keV ($\rm O^{7+}$), 
1.02 keV ($\rm Ne^{9+}$), 1.32 keV ($\rm Ne^{9+}$), 1.47 keV ($\rm Mg^{11+}$), 1.86 keV ($\rm Si^{12+}$), and 1.97 keV ($\rm Si^{13+}$) 
are typically seen in the observed geocoronal SWCX spectrum. 
Specially, we observed an unusual emission line at about 0.53 keV. 
It is most likely the $\rm N^{5+}$ $1s^1 5p^1 \to 1s^2$ transition which provides a strong evidence that 
the excess emission is of a non-thermal origin.
Another possibility is that there is fluorescent emission of $\rm O$ K$\alpha$ line at 0.525 keV reflected 
inside the Earth atmosphere. However, we do not observe comparable amount of $\rm N$ K$\alpha$ line at 0.392 keV which 
is another fluorescent line expected to emerge as seen in the Earth albedo spectrum. Therefore, 
the 0.53 keV line is not very likely to arise from the Earth atmosphere fluorescence.
Besides, the emission around $\sim$0.8 keV has a broad energy distribution and has been usually 
recognized as the Fe-L shell complex. The fit with CX models, however, suggested that this bump 
at $\sim$0.8 keV is possibly contributed by oxygen (see Figure \ref{fig:swcx_gaussian} (b)-1 and (c)-1).
Another point worthy of discussion is the 1.32 keV $\rm Ne^{9+}$ line, around this energy the 
line could also be recognized as 1.329 keV $\rm Mg^{10+}$, whereas the CX model favors to fit the 
1.32 keV line with $\rm Ne^{9+}$. 

In Table \ref{table:gaussian_fit} we show the best-fitted 
temperature and metal abundances with the CX model in SPEX and XSPEC. Both models are based on 
the collisional ionization equilibrium assumption, which may not hold for 
rapidly expanding CME plasmas that have not reached the equilibrium state when becoming collisionless. 
The fit results show significant degeneracies among the temperature, velocity, and metal abundances 
in SPEX-CX model. The XSPEC-VACX2 model suggests a larger collisional velocity, or higher element 
abundance ratio to oxygen for heavier ions than SPEX-CX model. 


\begin{sidewaystable}

\small

\caption{Model fit parameters for SWCX spectrum}
\centering
\begin{tabular}{p{0.1\textwidth}ccccc}
\hline
\hline
 Model & Energy & Line Identification & Surface Brightness & Flux ratio to $\rm O^{7+}$  \\
\hline
 Gaussian & (eV) &   & ($\rm photons\ cm^{-2}\ s^{-1}\ str^{-1}$) &  \\
\hline
& $530_{-15}^{+ 4}$ & $\rm N^{5+}$ $1s^1 5p^1 \to 1s^2$ (533 eV) & $2.4_{-0.5}^{+0.5}$ & $0.42$ \\
& $647_{-4}^{+ 5}$ & $\rm O^{7+}$ $2p^1 \to 1s^1$ (653 eV) & $5.7_{-0.6}^{+0.6}$ & $1$ \\
& $813_{-29}^{+ 16}$ & $\rm Fe^{16+}$ $2p^5 3d^1 \to 2p^6$ (826 eV,812 eV) & $2.6_{-0.4}^{+0.4}$ & $0.46$ \\ 
& $1028_{-4}^{+ 5}$ & $\rm Ne^{9+}$ $2p^1 \to 1s^1$ (1.022 keV) & $1.39_{-0.18}^{+0.18}$ & $0.24$ \\
& $1089_{-87}^{+ 67}$ & $\rm Fe^{23+}$ $1s^2 3s^1 \to 1s^2 2p^1$ (1.085 keV), $\rm Fe^{17+}$, $\rm Fe^{21+}$ & $1.1_{-0.2}^{+0.2}$ & $0.19$  \\
& $1320_{-10}^{+ 9}$ & $\rm Ne^{9+}$ $6p^1 \to 1s^1$ (1.324 keV) & $0.57_{-0.10}^{+0.10}$ & $0.10$  \\
& $1460_{-5}^{+ 6}$ & $\rm Mg^{11+}$ $2p^1 \to 1s^1$ (1.472 keV) & $0.8_{-0.1}^{+0.1}$ & $0.14$  \\
& $1857_{-22}^{+ 23}$ & $\rm Si^{12+}$ $1s^1 2p^1 \to 1s^2$ (1.865 keV) & $0.6_{-0.2}^{+0.2}$ & $0.10$  \\
& $1973_{-36}^{+ 26}$ & $\rm Si^{13+}$ $2p^1 \to 1s^1$ (2.006 keV) & $0.2_{-0.1}^{+0.1}$ & $0.035$  \\
\hline
\\
Model     &  kT (keV) &  Abundance   & Velocity ($\rm km \ s^{-1}$) & $\chi^2 / d.o.f$ \\
\\
\hline
CX   &   $1.04_{-0.01}^{+0.01}$      &    Solar $\rm abundance^{\dagger}$      &     $592_{-56}^{+113}$       &  4.54  \\ 
(SPEX)          &   $1.14_{-0.02}^{+0.02}$  &    $\rm C/C_{\odot}=8.2_{-0.9}^{+0.9}$ $\rm N/N_{\odot}=6.7_{-0.6}^{+0.7}$       &     $232_{-25}^{+45}$   &  4.16  \\
			&							&	$\rm Ne/Ne_{\odot}=0.8_{-0.05}^{+0.05}$  $\rm Mg/Mg_{\odot}=0.91_{-0.07}^{+0.08}$ $\rm Si/Si_{\odot}=0.43_{-0.06}^{+0.07}$		&  &  \\						
VACX2    &  1.16  &    Solar abundance  &  448.617  &  3.62  \\
(XSPEC)  &  1.05  &    $\rm C/C_{\odot}=3.67245$ $\rm N/N_{\odot}=5.26688$   &   854.947  & 3.22  \\
     &		  &    $\rm Ne/Ne_{\odot}=1.90699$ $\rm Mg/Mg_{\odot}=4.15255$ $\rm Si/Si_{\odot}=1.78754$  &    &  \\
\hline
\multicolumn{5}{l}{ $^\dagger$ Solar abundance refers to the abundance in the solar photosphere \cite{Anders_Grevesse1989}.}
\end{tabular}
\label{table:gaussian_fit}
\end{sidewaystable}

\section{SWCX as a result of the interaction between the interplanetary and Earth magnetic field}
\label{section-3}

We found two observational evidences suggesting that the observed SWCX emission is associated 
with the interplanetary and Earth magnetic field: (1) A time coincidence between the arrival of 
the magnetic cloud (MC) and the SWCX occurence. (2) An orbital modulation in SWCX light curve. 

Figure \ref{fig:region_lc} shows the light curves of the soft X-ray and proton flux detected 
by GOES, and magnetic field strength, alpha to proton density ratio, and proton density 
measured in-situ by ACE, together with the SWCX emission detected by Suzaku. 
The SWCX flaring starts around 2006-Dec-15th 00:00 UT, which is right after the ACE magnetic 
field strength entering a smooth transition phase, indicating the passage of an MC \cite{Liu2008}.
Before the MC phase, ACE also detected significant fluctuation in the field strength and the proton density between 
14:00 to 24:00 UT on 2006-Dec-14th, corresponding to the turbulent sheath of the CME shock front.
This is consistent with the CME propagation in the standard flare model illustrated in Figure \ref{fig:schematic},
in which the magnetic reconnection between the coronal loops launches a CME with a magnetic flux rope 
as its skeleton. When arriving at the Earth, the flux rope and Earth field could reconnect 
so that the heavily ionized ions are streamed into the magnetosphere, particularly to the polar 
regions, along the magnetic field lines.

Furthermore, we find that the X-ray background intensity has a significant time variation 
associated with the satellite orbital motion during the SWCX occurence period, shown in Figure \ref{fig:swcx_geometry} (a).
We reason that this time variation indicates that the source emission region is located close 
to the observer. Assuming the source distance is $D$, the maximum physics size the telescope 
can observe within the field of view (FoV with angular size $\theta$) is $D\theta$. Having witnessed 
the time variation of the diffuse emission suggests the source moves in and out of the telescope
FoV as satellite orbits around and the source region extension must be smaller than 
the orbit diameter $2R_{\rm orb}$. The distance is thus constrained to be $D < 2R_{\rm orb}/\theta = 370 R_{\rm E}$,
where $R_{\rm E}$ is the diameter of the Earth on the equator plane. 
Given that the observation was pointed towards the direction of the North pole,
it is likely that the line of sight passed through the cusp region of the Earth's magnetic field.  
We define the $r_{\rm mp}$ as the geocentric distance of the point where geomagnetic field becomes 
open to space for the first time along the line of sight starting from the spacecraft, and estimate $r_{\rm mp}$ 
using Tsyganenko geomagnetic field (T96) model \cite{Tsyganenko1995, Tsyganenko2005} 
and geopack libraries. Figure \ref{fig:sketch} shows a sketch of the $r_{\rm mp}$ definition related to the magnetosphere geometry.
The solar wind ram pressure is chosen to be $4.3\rm\ nPa$, with the proton density $3.59 \rm\ cm^{-3}$ 
and proton velocity $846.52\rm\ km\ s^{-1}$ as measured by ACE, and the interplanetary magnetic field 
strength of $B_y \sim 11\rm\ nT$ and $B_z \sim -13\rm\ nT$, and $Dst$ index $\sim -153\rm\ nT$
adopted as the input parameters of the model. 

As shown in Figure \ref{fig:swcx_geometry} (a), $r_{\rm mp}$ is in the range of 1.5 to $\sim 3 R_{\rm E}$, 
which is much smaller than the outer boundary of Earth magnetosphere, 10 $R_{\rm E}$, throughout the 
whole orbit period, suggesting the satellite line of sight passed through the cusp region. 
The strong SWCX emission could be explained by the penetration of solar-wind ions to the cusp region 
\cite{Walsh2016} where the neutral hydrogen density is high, with the solar wind being resisted 
to penetrate into any deeper region below $r_{\rm mp}$ due to the strong magnetic pressure.
If the neutral hydrogen density is the dominant factor determining the SWCX intensity, the maximum emission 
is expected at the lowest $r_{\rm mp}$ point, which is, however, not the case. 
On the other hand, variation in solar wind ion density dominated by gravity-pressure 
balance in the ionosphere is less likely to account for the SWCX intensity 
orbital dependence, either, since the solar wind will stop falling below the ionosphere boundary $\sim 2.5 R_{\rm E}$. 
Therefore, radial variation in neutral hydrogen or solar wind ion density cannot account 
for the observed light curve and another mechanism is necessary. Azimuthal solar wind ion intensity 
non-uniformity inside the cusp region provides one possible solution. 
In Figure \ref{fig:swcx_geometry} (b), we plot the X-ray intensity and $r_{\rm mp}$ as functions 
of the azimuthal angle of the $r_{\rm mp}$ position in GSM (geocentric solar magnetospheric coordinates) 
for the second orbital cycle, showing that the SWCX intensity is high when the satellite line of sight 
passes through the dusk side of the cusp, which means an azimuthal asymmetry exist in the solar wind density 
distribution in the cusp. Such density gradient may be related to the Parker spiral-type propagation 
of solar wind enhancement in the planetary space \cite{Koutroumpa2007}  
in which scenario the solar wind enhancement due to a CME event first arrives from the dusk side of the Earth.  
However, the azimuthal angle of the hit point in GSM is likely $< 90^\circ$, 
while the peak of the X-ray intensity is at $\sim 130^\circ$. 
We consider it is necessary to model solar wind transportation with azimuthal non-uniformity in the cusp.   

\section{Modelling the orbital modulation of the SWCX count rate}
\label{section:4}

The surface brightness of SWCX, $S$ is estimated by integrating the emission intensity per volume along the line of sight $s$,
\begin{equation}
S = \frac{1}{4\pi}\int_{s_0}^{s_1}{ \alpha n_{\rm H}(s) n_{\rm ion}(s) v_{\rm col} ds}.
\end{equation}
Then the observed intensity, $R$, is obtained by convolving $S$ with the telescope and detector response functions, and is approximately, 
$R = A\Omega S$, where $A$ and $\Omega$ are, respectively, the effective area and the field of view of the instrument. 
For CX emission line of certain ions, for example, $O^{8+}$, the ion density is estimated by solar wind ion 
density and ionic abundance, $n_{\rm ion} = n_{sw}  \chi_{O^{8+}}$.

$v_{\rm col}$ is the relative collisional speed between solar wind ions and neutral atoms in the Earth atmosphere.
Here we assume $v_{\rm col} = 500\rm\ km\ s^{-1}$, a speed smaller than the 1 AU speed of the CME shock, $\sim$1030 $\rm km\ s^{-1}$ \cite{Liu2008}, 
and neglect the thermal velocity.
The charge exchange cross section, $\alpha$, is chosen as $5\times 10^{-15}\rm\ cm^2$ for $O^{8+}+H$ collisions at 
$v_{\rm col} = 500\rm\ km\ s^{-1}$ \cite{Lee2004,Kimura1987}.

The neutral hydrogen density along line-of-sight $n_{\rm H}(s)$ is a function of the position in the Earth coordinate,
which, more specifically, is a function the distance from the Earth center, i.e. $n_{\rm H}(s) = n_{\rm H}(r(s))$. 
We use the geocoronal model proposed by \citeA{OStgaard2003} as an approximation to the 
neutral hydrogen density distribution in the Earth atmosphere, $n_{\rm H}(r) = 10000 e^{-r / 1.02}+ 70 e^{-r / 8.2}$, 
where $r$ is in the unit of $R_{\rm E}$ and $n_{\rm H}(r)$ in the unit of $\rm cm^{-3}$.
The model is based on the measurements of the Lyman $\alpha$ column brightness by the 
Geocoronal Imager on board the IMAGE satellite at high altitudes ($>$ 3.5 $R_{\rm E}$) where the 
neutral hydrogen is considered as an optical thin medium. Our estimation of neutral hydrogen 
density may suffer from large uncertainty due to the following reasons: (1) The variation of solar Lyman $\alpha$ flux 
itself from solar minimum to solar maximum is about a factor of 1.5 on average to 2.1 maximum; 
(2) We extrapolate the empirical model to smaller altitudes where the medium is no longer optically 
thin and more complex analysis including radiative transfer is required; (3) Asymmetric exosphere, 
"magnetotail", may present higher densities at high altitudes in the night-side than day-side direction 
under the radiation pressure.

The solar-wind ion density, $n_{\rm sw}(s)$, is also a function of the position in the Earth coordinate.  
\citeA{Walsh2016} showed the ion density as a function of the distance from the Earth center using numerical simulations.  
Their result shows that the density drops from  $\sim 9 R_{\rm E}$ to $\sim 6 R_{\rm E}$, 
and stays constant down to $\sim 2.5 R_{\rm E}$, then becomes zero. 
According to ACE satellite we found the proton density of the CME shock peaked around 15 $\rm cm^{-3}$.
We thus normalize the $n_{\rm sw}(s)$ at the outer boundary of the magnetosphere to be 15 $\rm cm^{-3}$ 
and use the radial profile in \citeA{Walsh2016} as the solar-wind ion density radial dependence along line of sight in our model.
As discussed in the previous section, 
we need to introduce density gradient from sun side to anti-sun side with in the cusp.  
We have no physical model for this, though. We thus assume a simple toy model, in which the 
density is also a function of $\Theta$ defined in the previous section, 
$n_{\rm ion}(s) = n_{\rm ion}(r(s), \Theta(s))$.  
The solar-wind ion density orbital modulation function is defined as $f(\Theta) = M e^{-\frac{1}{2} \frac{(\Theta - \mu)^2}{\sigma^2}}$,
where $\mu = 0.5, \sigma = 0.5$, and M is a normalization factor. $\mu$ is chosen to be 0.5 such that 
the solar-wind density is the highest when the satellite moves closest to the sun.
The abundance of $O^{8+}$ ions is estimated as its photospheric value, $8.51\times 10^{-4}$ \cite{Anders_Grevesse1989}.

With the assumptions discussed above, the model predicts  $\rm O^{7+}$ the count rate to be 0.01 $\rm cts\ s^{-1}$.  
On the other hand, from the energy spectrum, we estimate the observed $\rm O^{7+}$ count rate at the peak of 
figure \ref{fig:swcx_geometry} (a) to be $\sim 0.5\ {\rm cts\ s^{-1}}$, which exceeds the 
model predication by a factor about 50.
This large discrepancy could be related the non-uniformity of the flow in the cusp region 
as discussed in section \ref{section-3}. If the plasma flow is confined in 1/50 of the whole cusp volume,
the density is likely enhaced by this factor compared to the prediction with the uniform model from \citeA{Walsh2016}. 
In figure \ref{fig:swcx_geometry} (b), the FWHM of the peak is about 0.8 radian, which is about 1/8 of $2\pi$.  
A factor of 50 times change in volume can be explained if the plasma is compressed in two directions with this ratio.

\section{Summary and conclusion}
\label{section:5}

We report an unique case of SWCX event caused by the well-known 2006-December-13 CME, which 
is captured by Suzaku observation with a line-of-sight cutting through the cusp region. 
Significant non-thermal emission lines from $\rm O^{7+}$, $\rm Ne^{9+}$, $\rm Mg^{11+}$, $\rm Si^{12+}$, 
$\rm Si^{13+}$, and probably, $\rm N^{5+}$ $1s^1 5p^1 \to 1s^2$ transition are detected and we fit the SWCX spectrum 
with CX models. We find a time coincidence between the arrival of CME magnetic cloud and the SWCX occurrence, suggesting 
that the solar wind ions can be streamed onto the Earth field lines via magnetic reconnection.  
The time variation of the X-ray emission suggests non-uniformity of the solar wind flow in the cusp region 
which enhances the X-ray surface brightness by a factor of $\sim 50$.  
The SWCX is bright in the dusk side of the cusp and we conjecture that this is related to the 
Parker spiral-type propagation of the solar wind.

%
%

\section*{Open Research Section}
Suzaku archive data available from the DARTS/Suzaku Public Data list 
(\url{https://data.darts.isas.jaxa.jp/pub/Astro_Browse/publications/suzaku/public_list.html}) 
were used in the creation of this manuscript. 
The Tsyganenko Geomagnetic Field Model and GEOPACK libraries are published on Github 
\url{https://github.com/tsssss/geopack}.

\acknowledgments
We thank professor Hiroya Yamaguchi and Yuichiro Ezoe for insightful discussions and KAKENHI grants: JP20KK0072 (PI: S. Toriumi), JP21H01124 (PI: T. Yokoyama), and JP21H04492 (PI: K. Kusano).

\bibliography{refs}

\end{document}